\documentstyle[preprint,aps,version2]{revtex}
\draft
\def\Journal#1#2#3#4{{#1} {#2} {(#4)} {#3}}

\def\NP{{ Nucl. Phys.} }
\def\PLB{{ Phys. Lett.}  B}
\def\PL{{ Phys. Lett.}}
\def\PR{{ Phys.   Rep.}}
\def\PRL{ Phys. Rev. Lett.}
\def\PRD{{ Phys. Rev.} D}
\def\PRC{{ Phys. Rev.} C}
\def\ZPC{{Z. Phys.} C}
\def\EPJC{{Eur. Phys. J.} C}

\def\MPLA{{Mod. Phys. Lett.} A}
\def\CPC{Comput. Phys. Commun.}

\def\ra{\rightarrow}

\def\be{\begin{equation}}
\def\ee{\end{equation}}
\def\bea{\begin{eqnarray}}
\def\eea{\end{eqnarray}}

\def\qbar{{\bar q}}
\def\ubar{{\bar u}}
\def\dbar{{\bar d}}
\def\sbar{{\bar s}}

\def\NP{{ Nucl. Phys.}}
\def\ANP{{Adv. Nucl. Phys.}}

\begin{document}
\begin{titlepage} 

\bigskip
\begin{center}
{\large\bf 
Charge symmetry violation in the parton distributions \\ of the nucleon}

\author{Fu-Guang Cao\thanks{E-mail address: f.g.cao@massey.ac.nz.}
 and A. I. Signal\thanks{E-mail address: a.i.signal@massey.ac.nz.}}
\begin{instit}
Institute of Fundamental Sciences \\ Massey University \\
Private Bag 11 222,  Palmerston North \\
New Zealand
\end{instit}
\end{center}

\begin{abstract}

We point out that charge symmetry violation in both the valence
and sea quark distributions of the nucleon has a non-perturbative source.
We calculate this non-perturbative charge symmetry violation using the 
meson cloud model, which has earlier been successfully applied to both the 
study of SU(2) flavour asymmetry in the nucleon sea and quark-antiquark 
asymmetry in the nucleon.
We find that the charge symmetry violation in the valence quark distribution
is well below $1\%$, which is consistent with most low energy tests
but significantly smaller than the quark model prediction about $5\%-10\%$.
Our prediction for the charge symmetry violation in the sea quark distribution
is also much smaller than the quark model calculation.

\bigskip
\vskip 1cm
\noindent
PACS numbers: 12.39.-x; 11.30.Er

\noindent
Keywords: Charge symmetry, Meson cloud, Parton distribution
\end{abstract}
\end{titlepage}

\section{Introduction}

Recently charge symmetry violation (CSV) in the parton distributions of the 
nucleon has attracted great interest.
It has been generally believed that charge symmetry (CS)
was highly respected in the nucleon system.
Most low energy experiments have shown that CS is satisfied within
about $1\%$ in reaction amplitudes \cite{LowEnergyTest},
and most high energy tests are also consistent with the charge symmetry,
although generally with less precision than the low energy tests (for a recent
review see \cite{LonderganT} and references therein).
However some theoretical calculations have suggested that the CSV
in the valence quark distributions may be as large as
$5\% - 10\%$ \cite{Sather,RodionovTL,BeneshG,BeneshGS,BeneshL,Ma}
which is rather large compared with the low-energy results.
There have been proposed a number of experiments in which CSV may be 
observed \cite{Measurements}.

A serious challenge to CS has come from the comparison of
the $F_2$ structure functions measured in charged and uncharged lepton 
deep inelastic scattering \cite{NMC,CCFR}
performed by  Boros, Londergan and Thomas \cite{BorosLT}.
A significantly larger CSV than the expectations of both theory
\cite{Sather,RodionovTL,BeneshG,BeneshGS,BeneshL,Ma}
and other experiments \cite{Other} was found for the $s$ and $\sbar$
distributions in the low-$x$ region ($x < 0.1$).
Any unexpected large CSV will greatly affect our understanding
of non-perturbative dynamics and hadronic structure \cite{LonderganT},
and also the extraction of $sin^2\theta_W$ from neutrino scattering \cite{Sather}.
More recently, Boros, Steffens, Londergan and Thomas \cite{BorosSLT}
performed a similar analysis to Ref.~\cite{BorosLT} with improved corrections
for nuclear shadowing and the charm threshold in the neutrino data.
They found that the data (including the low-$x$ region where a large
discrepancy was found in \cite{BorosLT})
are consistent with charge symmetry within experimental errors
and the present uncertainty in the strange quark distribution of the nucleon.

Thus suggestions of any large CSV in the parton distributions of the nucleon
result from theoretical calculations
\cite{Sather,RodionovTL,BeneshG,BeneshGS,BeneshL,Ma}.
Most of these theoretical calculations are performed using a quark model, 
such as the MIT bag model \cite{RodionovTL},
Los Alamos Potential Model \cite{BeneshG,BeneshGS},
or a model independent version \cite{Sather,BeneshL}.
The quark model calculations are based on a quark-spectator
(quark-diquark) picture of the nucleon in deep inelastic reactions,
which is questionable in the low-$x$ region.
Hence the quark model predictions for the low-$x$ behaviour of CSV
are not very reliable.
For the sea quark content, quark model calculations involve spectator states
containing four quarks, which is also unreliable in the low-$x$ region as the 
mass parameter for these four-quark (or three-quark one-antiquark) states
is less well-determined than that for the diquark states.
CSV has also been estimated using the light-cone baryon-meson
fluctuation model \cite{Ma}. However, this calculation is mostly qualitative --
the quantitative calculations are highly dependant on the model parameters.
More theoretical study on CSV from a
different point of view than the quark model will be worthwhile.

In this paper, we point out that
the CSV in both the valence quark and sea quark has the same
non-perturbative source, and that
the meson cloud model (MCM) -- which has been successfully applied to
the study of the sea quark content of nucleon (including SU(2) flavour asymmetry
\cite{Thomas83,Holtmann,Speth,MelnitchoukST,Kumano},
and $s$ and $\sbar$ content of the nucleon \cite{Holtmann,SignalT,CaoS}) --
can provide a natural explanation of CSV in the
valence quark and sea quark distributions of the nucleon.
We shall make an alternative investigation
of CSV in the parton distributions of the nucleon
by using the meson cloud model instead of the more commonly used quark model.
Our calculations for the CSV are significantly different from
the quark model predictions.

\section{Charge symmetry violation in the meson cloud model}

Charge symmetry results from the $180^o$-rotation invariance of the strong
Hamiltonian about the $2$-axis in isospace \cite{LonderganT}.
At the quark level charge symmetry implies the invariance of a system
under the interchange of up and down quarks.
For the valence and sea parton distributions, this results in the following relations:
\bea
u^p_v(x)=d^n_v(x), &~~~&
d^p_v(x)=u^n_v(x), \\
\label{ValanceCS}
\ubar^p(x)=\dbar^n(x), &~~~&
\dbar^p(x)=\ubar^n(x), \\
s^p(x)=s^n(x), &~~~&
\sbar^p(x)=\sbar^n(x).
\label{SeaCS}
\eea
The charge symmetry violation in the parton distributions of the nucleon
can be `measured' via the quantities:
\bea
\delta d_v(x)=d^p_v(x)-u^n_v(x), &~~~&
\delta u_v(x)=u^p_v(x)-d^n_v(x), \\
\delta \dbar(x)=\dbar^p(x)-\ubar^n(x), &~~~&
\delta \ubar(x)=\ubar^p(x)-\dbar^n(x), \\
\delta s(x)=s^p(x)-s^n(x), &~~~&
\delta \sbar(x)=\sbar^p(x)-\sbar^n(x).
\label{CSV}
\eea


Before entering detailed calculation, it is helpful to break down the parton
distribution in the nucleon into three parts:
the parton distribution in the bare nucleon, the perturbative contribution,
and the non-perturbative contribution {\it i.e.}
\bea
d^p = d^p_{\rm bare}+d^p_{\rm per}+d^p_{\rm non}, &~~&
u^p = u^p_{\rm bare}+u^p_{\rm per}+u^p_{\rm non}, \\
\dbar^p = \dbar^p_{\rm per}+\dbar^p_{\rm non}, &~~&
\ubar^p = \ubar^p_{\rm per}+\ubar^p_{\rm non}, \\
s^p = s^p_{\rm per}+s^p_{\rm non}, &~~&
\sbar^p = \sbar^p_{\rm per}+\sbar^p_{\rm non}.
\label{Classfication}
\eea
Similar relations exist for the parton distribution of the neutron.
We expect that the bare part obeys the charge symmetry
\bea
d_{\rm bare}^p=u^n_{\rm bare},
~~~
u_{\rm bare}^p=d^n_{\rm bare}.
\label{Relation_bare}
\eea
The perturbative sea is produced in a very short time
via gluon splitting, thus we expect the perturbative sea to also be
SU(2) flavour symmetric,
\bea
\dbar^p_{\rm per}=\ubar^p_{\rm per},
~~~
\ubar^p_{\rm per}=\dbar^n_{\rm per},
\label{flavour_symmetric}
\eea
quark-antiquark symmetric,
\bea
q^{p,n}_{\rm per}=\qbar^{p,n}_{\rm per}, ~~~ (q=u,d,s)
\label{qqbar_symmetric}
\eea
and charge symmetric,
\bea
d^p_{\rm per}=u^n_{\rm per},
&~~~&
u^p_{\rm per}=d^n_{\rm per},
\label{CS1}\\
\dbar^p_{\rm per}=\ubar^n_{\rm per},
&~~~&
\ubar^p_{\rm per}=\dbar^n_{\rm per},
\label{CS2}\\
s^p_{\rm per}=s^n_{\rm per},
&~~~&
\sbar^p_{\rm per}=\sbar^n_{\rm per}.
\label{CS3}
\eea
From the bare parton distribution being charge symmetric [Eq.~(\ref{Relation_bare})]
and the perturbative sea being quark-antiquark symmetric
[Eq.~(\ref{qqbar_symmetric})] and charge symmetric
[Eqs.~(\ref{CS1}) and (\ref{CS2})] we have the CS violating valence distributions
\bea
\delta d_v
 &=& (d^p_{\rm non}-\dbar^p_{\rm non})
 -(u^n_{\rm non}-\ubar^n_{\rm non}),
\label{delta_dv} \\
\delta u_v
 &=& (u^p_{\rm non}-\ubar^p_{\rm non})
 -(d^n_{\rm non}-\dbar^n_{\rm non}).
\label{delta_uv}
\eea
Using the charge symmetry of the perturbative sea  
[Eqs.~(\ref{CS1})-(\ref{CS3})] we can obtain the CS violating sea distributions
\bea
\delta \dbar =\dbar^p_{\rm non}-\ubar^n_{\rm non},
&~~~&
\delta \ubar =\ubar^p_{\rm non}-\dbar^n_{\rm non},
\label{delta_dubar} \\
\delta s=s^p_{\rm non}-s^n_{\rm non}, 
&~~~&
\delta \sbar=\sbar^p_{\rm non}-\sbar^n_{\rm non}.
\label{delta_ssbar}
\eea
Thus the charge symmetry violation in both the valence 
and the sea distributions has a non-perturbative origin.

The meson cloud model (MCM) is a model of the non-perturbative
contribution to the quark distributions of the nucleon.
It can provide natural explanations of the flavour
asymmetry in the nucleon sea
\cite{Thomas83,Holtmann,Speth,MelnitchoukST,Kumano}
and quark-antiquark asymmetry \cite{Holtmann,SignalT,CaoS}
in the nucleon.
The essential point of the MCM is that the nucleon can fluctuate into
different baryon-meson Fock states,
\bea
|N\rangle_{\rm physical} =  Z |N\rangle_{\rm bare}
+\sum_{BM} \sum_{\lambda \lambda^\prime} 
\int dy \, d^2 {\bf k}_\perp \, \phi^{\lambda \lambda^\prime}_{BM}(y,k_\perp^2)
\, |B^\lambda(y, {\bf k}_\perp); M^{\lambda^\prime}(1-y,-{\bf k}_\perp)
\rangle 
\label{NMCM}
\eea
where $Z$ is the wave function renormalization constant,
$\phi^{\lambda \lambda^\prime}_{BM}(y,k_\perp^2)$ 
is the wave function of the Fock state containing a baryon ($B$)
with longitudinal momentum fraction $y$, transverse momentum ${\bf k}_\perp$,
and helicity $\lambda$,
and a meson ($M$) with momentum fraction $1-y$,
transverse momentum $-{\bf k}_\perp$, and helicity $\lambda^\prime$.
The model assumes that the lifetime of a virtual baryon-meson Fock state is much
larger than the interaction time in the deep inelastic or Drell-Yan
process, thus the contribution from the virtual baryon-meson Fock states to
the quark and anti-quark distributions of the nucleon can be written as convolutions
\bea
q_{\rm non}(x)&=&\sum_{BM} \left[\int^1_x \frac{dy}{y} f_{BM/N}(y) q^B(\frac{x}{y})
		   +\int^1_x \frac{dy}{y} f_{MB/N}(1-y) q^M(\frac{x}{y}) \right],
\label{qBM} \\
\qbar_{\rm non}(x)&=&
\sum_{BM} \int^1_x \frac{dy}{y} f_{MB/N}(1-y) \qbar^M(\frac{x}{y}),
\label{qbarBM}
\eea
where $f_{B M/N}(y)=f_{MB/N}(1-y)$ is fluctuation function which gives
the probability for the nucleon fluctuating into a virtual $BM$ state
\bea
f_{BM/N}(y)=\int^\infty_0 d k_\perp^2 \left | \phi_{B M}(y, k_\perp^2)\right |^2.
\label{fBM}
\eea
As the proton and neutron form an SU(2) isospin doublet,
the baryons and mesons in their respective virtual Fock states differ
only in their carried charge.
If we neglect the mass differences among these baryons and mesons
the fluctuation functions for the proton and neutron will be the same.
Thus the contributions to the parton distribution of the nucleon
from these fluctuations will be isospin symmetric (charge symmetric).
As is well known, the electromagnetic interaction induces mass differences
among these baryons and mesons. If we take into account these mass
differences, the probabilities for the corresponding fluctuations
of proton and neutron will be different and
thus the contributions to the parton distributions of the proton
and neutron will be different, which results in  CSV
in the parton distributions of the nucleon.
Thus the MCM can provide a natural explanation of CSV
in the parton distributions of the nucleon.
Although it is argued from the quark model calculations
that the electromagnetic effect does not play a significant role
in the calculation of CSV in the parton distributions,
it is worthwhile to study this effect using a different theoretical picture.

For the CSV in the up and down quark distributions,
we consider the fluctuations $N \ra N \pi$ and $N \ra \Delta \pi$,
but neglect the other fluctuations such as $N \ra N \, (\Delta) \rho$,
$N \ra N \eta \, (\omega)$ and $N \ra \Delta \eta \, (\omega)$
due to the higher masses of the involved mesons.
Thus the fluctuations we consider include:
\bea
p(uud) \ra n(udd) + \pi^+(u\dbar), &~~~&
n(udd) \ra p(uud) + \pi^-(\ubar d), \nonumber \\
p(uud) \ra \Delta^0(udd) + \pi^+(u\dbar), &~~~&
n(udd) \ra \Delta^+(uud) + \pi^-(\ubar d), \nonumber \\
p(uud) \ra p(uud) + \pi^0([u \ubar + d \dbar]/\sqrt{2}), &~~~&
n(udd) \ra n(udd) + \pi^0([u \ubar + d \dbar]/\sqrt{2}), \nonumber \\
p(uud) \ra \Delta^+(uud) + \pi^0([u \ubar + d \dbar]/\sqrt{2}), &~~~&
n(udd) \ra \Delta^0(udd) + \pi^0([u \ubar + d \dbar]/\sqrt{2}), \nonumber \\
p(uud) \ra \Delta^{++}(uuu) + \pi^-(\ubar d), &~~~&
n(udd) \ra \Delta^{-}(ddd) + \pi^+(u \dbar).
\label{fluctuation1}
\eea
For the CSV in the strange and anti-strange quark
distributions we consider the following fluctuations
\bea
p(uud) \ra \Lambda(uds) + K^+(u \sbar), &~~~&
n(udd) \ra \Lambda(uds) + K^0(d \sbar), \nonumber \\
p(uud) \ra \Sigma^0(uds) + K^+(u \sbar), &~~~&
n(udd) \ra \Sigma^0(uds) + K^0(d \sbar), \nonumber \\
p(uud) \ra \Sigma^+(uus) + K^0(d \sbar), &~~~&
n(udd) \ra \Sigma^-(dds) + K^+(u \sbar).
\label{fluctuation2}
\eea
Since we consider the CSV between the proton and neutron, we should not
neglect the mass difference between the proton and neutron, 
$m_p-m_n=-1.3 \, {\rm MeV}$, and those among the baryon and
meson multiplets \cite{PDG98},
\bea
m_{\Delta^-}-m_{\Delta^0} & = & m_{\Delta^0}-m_{\Delta^+} =
m_{\Delta^+}-m_{\Delta^{++}}  =  1.3 \, {\rm MeV}, \nonumber \\
m_{\Sigma^-}-m_{\Sigma^0} & = & 4.8 \, {\rm MeV}, \nonumber \\
m_{\Sigma^0}-m_{\Sigma^+} & = & 3.3 \, {\rm MeV}, \nonumber \\
m_{\pi^\pm}-m_{\pi^0} & = & 4.6 \, {\rm MeV}, \nonumber \\
m_{K^0}-m_{K^+} & = & 4.0 \, {\rm MeV}.
\eea

The probabilities of various fluctuations can be calculated using
the effective Lagrangian and time-ordered perturbation theory in the
infinite momentum frame.
For the fluctuations $N \ra N \pi$,
$N \ra \Lambda K$ and $ N \ra \Sigma K$,
the fluctuation functions can be expressed as \cite{Holtmann,MelnitchoukST}
\bea
f_{BM/N}(y)=C_1\frac{ g_{NBM}^2}{16 \pi^2}
\int^\infty_0 \frac{d k_\perp^2}{y (1-y)}
\frac{G^2_{BM}(M^2_{BM})}{(m^2_N-m^2_{BM})^2}
\frac{k_\perp^2+(y m_N -m_B)^2}{y},
\label{fBMN}
\eea
where $y$ is the longitudinal momentum fraction of the baryon $B$,
$g_{NBM}$ is the effective coupling constant,
$m_{BM}^2$ is the invariant mass squared of the $BM$ Fock state,
\bea
m_{BM}^2=\frac{m_B^2+k_\perp^2}{y}
+\frac{m_\pi^2+k_\perp^2}{1-y},
\label{mBM2}
\eea
and $G_{BM}$ is the phenomenological form factor, for which we adopt
the exponential form 
\bea
G_{BM}(y,k_\perp^2)={\rm exp}\left[
\frac{m_N^2-m_{BM}^2(y,k_\perp^2)}{2 \Lambda^2}\right].
\label{GBM}
\eea
$\Lambda$ is a cut-off parameter which can be taken
as $\Lambda=1.08 {\rm GeV}$ for all fluctuations involving
octet baryons and pseudoscalar or vector mesons \cite{Holtmann}.
The effective coupling constants are taken to be
$g_{NN\pi}=13.07$ \cite{Holtmann,MelnitchoukST},
$g_{N\Lambda K}=13.12$ and $g_{N\Sigma K}=6.82$ \cite{SignalT,Aubert}.
The coefficient $C_1$ in Eq.~(\ref{fBMN}) comes from the Clebsch-Gordan
coefficients for the fluctuations of different isospin multiplets.
$C_1=1$ for $ f_{p\pi^0/p}$, $f_{n\pi^0/n}$, $f_{\Lambda K^+/p}$,
$f_{\Lambda K^0/n}$, $f_{\Sigma^0 K^+/p}$ and $f_{\Sigma^- K^0/n}$,
and $C_1=2$ for $f_{n\pi^+/p}$, $f_{p\pi^-/n}$, $f_{\Sigma^+ K^0/p}$,
and $f_{\Sigma^- K^+/n}$.
For the fluctuation $N \ra \Delta \pi$ we have \cite{Holtmann,MelnitchoukST}
\bea
f_{\Delta \pi /N}(y)=C_2\frac{ g^2}{16 \pi^2}
\int^\infty_0 \frac{d k_\perp^2}{y (1-y)}
\frac{G^2_{\Delta \pi}(M^2_{\Delta \pi})}{(m^2_N-m^2_{\Delta \pi})^2}
\frac{[k_\perp^2+(m_\Delta - y m_N)^2]
[k_\perp^2+(m_\Delta+ym_N)^2]^2}
{6 m_\Delta^2 y^3},
\label{fDeltapiN}
\eea
where $g=11.8 \, {\rm GeV}^{-1}$ \cite{Holtmann,MelnitchoukST},
$C_2=1$ for $f_{\Delta^{++}\pi^-/p}$ and $f_{\Delta^{-}\pi^+/n}$,
$C_2=2/3$ for $f_{\Delta^{+}\pi^0/p}$ and $f_{\Delta^{0}\pi^0/n}$,
and $C_2
=1/3$ for $f_{\Delta^{0}\pi^+/p}$ and $f_{\Delta^{+}\pi^-/p}$.
We adopt the exponential form [Eq.~(\ref{GBM})] for the form factor
and the cut-off parameter was taken to be
$\Lambda=0.98 \, {\rm GeV}$ \cite{MelnitchoukST}.

In the meson cloud model the non-perturbative contribution
to the quark and the anti-quark distributions
in the nucleon sea come from the quarks and anti-quarks of the baryons
($N, \, \Lambda, \, \Sigma$) and mesons
($\pi, \, K$) in the virtual baryon-meson Fock states.
So we need the parton distributions in the involved baryons and mesons as input.
For the parton distribution in the pion, we employ the parameterization
given by Gl\"{u}ck, Reya, and Stratmann (GRS98) \cite{GRS98}
and we neglect the sea content in the meson, that is, 
\bea
\dbar^{\pi^+}=u^{\pi^+}&=&\ubar^{\pi^-}=d^{\pi^-}=\frac{1}{2} v^\pi, \\
\ubar^{\pi^0}=u^{\pi^0}&=&\dbar^{\pi^0}=d^{\pi^0}=\frac{1}{4} v^\pi, \\
v^\pi(x,\mu_{\rm NLO}^2) &=&1.052 x^{-0.495} (1 +0.357 \sqrt{x}) (1-x)^{0.365},
\label{vpion}
\eea
at scale $\mu_{\rm NLO}^2=0.34$ GeV$^2$.
For the $\sbar$ distribution in the $K^+$ and $K^0$ we use
the GRS98 parameterization \cite{GRS98}
\bea
\sbar^{K^+}(x,\mu_{\rm NLO}^2)=\sbar^{K^0}(x,\mu_{\rm NLO}^2)
=\left[1-0.540(1-x)^{0.17}\right] v^\pi(x,\mu_{\rm NLO}^2)
\label{sbar}
\eea
at scale $\mu_{\rm NLO}^2=0.34$ GeV$^2$.
For the quark distributions in the bare baryons, we first use
the up and down quark distributions in the proton
given by Gl\"{u}ck, Reya, and Vogt (GRV98) \cite{GRV98},
\bea
d^p(x,\mu_{\rm NLO}^2)&=&0.624 (1-x)u^p(x,\mu_{\rm NLO}^2) \\
u^p(x,\mu_{\rm NLO}^2)&=&0.632 \, x^{-0.57} (1-x)^{3.09} (1+18.2 x),
\eea
at scale $\mu_{\rm NLO}^2=0.40$ GeV$^2$, then relate these to the distributions 
in the other baryons via the relations
\bea
d^n=u^{\Delta^+}=d^{\Delta^0}=u^p, &~~~&
u^n=d^{\Delta^+}=u^{\Delta^0}=d^p, \\
u^{\Delta^{++}}=u^p+d^p, &~~~& d^{\Delta^-}=u^p+d^p, \\
s^\Lambda=s^{\Sigma^+}=s^{\Sigma^0}=s^{\Sigma^-} & = & \frac{1}{2}u^p. 
\label{PDinBaryon}
\eea
We evolve the distributions to the scale $Q^2 = 4$ GeV$^2$
using the program of Miyama and Kumano \cite{MiyamaK}
in which the evolution equation is solved numerically using a brute-force method.
We found that at $Q^2=4$ GeV$^2$ the parton distributions we need
($v^\pi(x, Q^2)$, $\sbar^{{\bar K^0}}(x, Q^2)$ $u^p(x, Q^2)$ and $d^p(x, Q^2)$)
can be parametrized using the following form
\bea
q(x, Q^2)=a \, x^b \, (1-x)^c\, (1+d \, \sqrt{x} +e \, x)
\label{qfit}
\eea
with the parameters given in Table 1.
We estimate the uncertainty in solving the evolution equations numerically
and parametrizating the parton distribution in the form of Eq.~(\ref{qfit})
to be about $2\%$ in the $x$-region which we are interested in
{\it ie} $x >10^{-3}$.

The final expressions for the CSV in the valence parton distributions
are given by:
\bea
x \delta d_v&=&\int_0^x d y \frac{x}{y}
\left\{
  \left[ f_{n \pi^+/p}(y)-f_{p \pi^-/n}(y) \right] u^p(\frac{x}{y}) \right. \nonumber \\
 & & ~~~~~~~\left. 
-\left[ f_{n \pi^+/p}(1-y)-f_{p \pi^-/n}(1-y) \right]
  \frac{1}{2} v^\pi(\frac{x}{y}) \right. \nonumber \\
& & ~~~~~~~\left.
+\left[ f_{p \pi^0/p}(y)-f_{n \pi^0/n}(y) \right] d^p(\frac{x}{y}) \right. \nonumber \\
& & ~~~~~~~\left.
+\left[ f_{\Delta^0 \pi^+/p}(y)-f_{\Delta^+ \pi^-/n}(y) \right] u^p(\frac{x}{y}) \right. \nonumber \\
& & ~~~~~~~\left.
-\left[ f_{\Delta^0 \pi^+/p}(1-y)-f_{\Delta^+ \pi^-/n}(1-y) \right]
  \frac{1}{2} v^\pi(\frac{x}{y}) \right. \nonumber \\
& & ~~~~~~~\left.
+\left[ f_{\Delta^+ \pi^0/p}(y)-f_{\Delta^0 \pi^0/n}(y) \right]
  d^p(\frac{x}{y}) \right. \nonumber \\
& & ~~~~~~~\left.
+\left[ f_{\Delta^{++} \pi^-/p}(1-y)-f_{\Delta^- \pi^+/n}(1-y) \right]
  \frac{1}{2} v^\pi(\frac{x}{y})
\right\},
\label{deltadv}
\eea
\bea
x \delta u_v&=&\int_0^x d y \frac{x}{y}
\left\{
  \left[ f_{n \pi^+/p}(y)-f_{p \pi^-/n}(y) \right] d^p(\frac{x}{y}) \right. \nonumber \\
& & ~~~~~~~\left.
-\left[ f_{n \pi^+/p}(1-y)-f_{p \pi^-/n}(1-y) \right]
  \frac{1}{2} v^\pi(\frac{x}{y}) \right. \nonumber \\
& & ~~~~~~~\left.
+\left[ f_{p \pi^0/p}(y)-f_{n \pi^0/n}(y) \right] u^p(\frac{x}{y}) \right. \nonumber \\
& & ~~~~~~~\left.
+\left[ f_{\Delta^0 \pi^+/p}(y)-f_{\Delta^+ \pi^-/n}(y) \right] d^p(\frac{x}{y}) \right. \nonumber \\
& & ~~~~~~~\left.
-\left[ f_{\Delta^0 \pi^+/p}(1-y)-f_{\Delta^+ \pi^-/n}(1-y) \right]
  \frac{1}{2} v^\pi(\frac{x}{y}) \right. \nonumber \\
& & ~~~~~~~\left.
+\left[ f_{\Delta^+ \pi^0/p}(y)-f_{\Delta^0 \pi^0/n}(y) \right]
  u^p(\frac{x}{y}) \right. \nonumber \\
& & ~~~~~~~\left.
+\left[ f_{\Delta^{++} \pi^-/p}(y)-f_{\Delta^- \pi^+/n}(y) \right]
  (u^p+d^p) \right. \nonumber \\
& & ~~~~~~~\left.
+\left[ f_{\Delta^{++} \pi^-/p}(1-y)-f_{\Delta^- \pi^+/n}(1-y) \right]
  \frac{1}{2} v^\pi(\frac{x}{y})
\right\}.
\label{deltauv}
\eea
For CSV in the sea we obtain
\bea
x \delta \dbar &=& \int_0^x d y \frac{x}{y}
\left\{
   f_{n \pi^+/p}(1-y)-f_{p \pi^-/n}(1-y)
  +\frac{1}{2} \left[ f_{p \pi^0/p}(1-y)-f_{n \pi^0/n}(1-y) \right] \right. \nonumber \\
& & ~~~~~~~~~~\left.
+ f_{\Delta^0 \pi^+/p}(1-y)-f_{\Delta^+ \pi^-/n}(1-y) \right. \nonumber \\
& & ~~~~~~~~~~\left.
   +\frac{1}{2} \left[ f_{\Delta^+ \pi^0/p}(1-y)-f_{\Delta^0 \pi^0/n}(1-y) \right]
\right\} \frac{1}{2} v^\pi(\frac{x}{y}),
\label{deltadbar}
\eea  
\bea
x \delta \ubar &=& \int_0^x d y \frac{x}{y}
\left\{
\frac{1}{2} \left[ f_{p \pi^0/p}(1-y)-f_{n \pi^0/n}(1-y) \right] 
  +\frac{1}{2} \left[ f_{\Delta^+ \pi^0/p}(1-y)-f_{\Delta^0 \pi^0/n}(1-y) \right]
  \right. \nonumber \\
& & ~~~~~~~~~~ \left.
  + f_{\Delta^{++} \pi^-/p}(1-y)-f_{\Delta^- \pi^+/n}(1-y)
  \right\} \frac{1}{2} v^\pi(\frac{x}{y}),
\label{deltaubar}
\eea  
\bea
x \delta s &=& \int_0^x d y \frac{x}{y}
\left\{
  f_{\Lambda K^+/p}(y)-f_{\Lambda K^0/n}(y)
+f_{\Sigma^0 K^+/p}(y)-f_{\Sigma^0 K^0/n}(y) \right. \nonumber \\
& &~~~~~~~~~~ \left.
+f_{\Sigma^+ K^0/p}(y)-f_{\Sigma^- K^+/n}(y)
\right\} \frac{1}{2} u^p(\frac{x}{y})
\label{deltas}
\eea
\bea
x \delta \sbar &=& \int_0^x d y\frac{x}{y}
\left\{
  f_{\Lambda K^+/p}(1-y)-f_{\Lambda K^0/n}(1-y)
+f_{\Sigma^0 K^+/p}(1-y)-f_{\Sigma^0 K^0/n}(1-y) \right. \nonumber \\
& &~~~~~~~~~~ \left.
+f_{\Sigma^+ K^0/p}(1-y)-f_{\Sigma^- K^+/n}(1-y)
\right\} \frac{1}{2} \sbar^K(\frac{x}{y})
\label{deltasbar}
\eea

\section{Result and discussion}

From Eqs.~(\ref{deltadv}) - (\ref{deltasbar})
we can see explicitly that the differences among various
fluctuation functions such as $f_{n\pi^+/p}$ and $f_{p\pi^-/n}$
result in the CSV in the parton distributions of the nucleon.
We plot these differences in Figs.~1-3.
It can be seen that the difference in the fluctuation $N\ra N \pi$
($f_{n\pi^+/p}-f_{p\pi^-/n}$, ...) is much larger than that in the
fluctuation $N \ra \Delta \pi$
($f_{\Delta^{++}\pi^-/p}-f_{\Delta^+\pi^+/n}$, ...)
and the latter is much bigger than that in the fluctuation $N \ra \Lambda K$
($f_{\Lambda\pi^+/p}-f_{\Lambda K^0/n}$, ...).
Thus the CSV in the valence and sea up and down quarks should be the same order,
and both larger than the CSV in the $s$ and $\sbar$ distributions.
The difference $f_{n\pi^+/p}-f_{p\pi^-/n}$ is much larger than
$f_{p\pi^0/p}-f_{n\pi^0/n}$, thus the CSV in the sea
of the minority quark flavor ($\delta \dbar$) will be much larger than
that of the majority quark flavor ($\delta \ubar$) due to the
absence of $(f_{n\pi^+/p}-f_{p\pi^-/n})$ term in $\delta \ubar$.
The probabilities of the various fluctuations can be obtained by integrating
the corresponding fluctuation functions.
We find the probabilities of the dominant fluctuations to be
\bea
P(p \ra n \pi^+)=0.202 &~~~& P(n \ra p \pi^-)=0.205, \\
P(p \ra \Delta^{++} \pi^-)=0.0481 &~~~& P(n \ra \Delta^- \pi^+)=0.0475, \\
P(p \ra \Lambda K^+)=0.0127 &~~~& P(n \ra \Lambda K^0)=0.0125,
\eea
that is there is about a $1\%$ excess of fluctuations
$n \ra p \pi^-$ over $p \ra n \pi^+$ and $p \ra \Delta^{++} \pi^-$ over
$n \ra \Delta^- \pi^+$, and about $2\%$ excess of $p \ra \Lambda K^+$
over $n \ra \Lambda K^0$.

We present our results for the CSV in the valence quark sector
($x \delta d_v$ and $x \delta u_v$) in Fig.~4.
We find that $x \delta d_v$ and $x \delta u_v$ have similar shape
and both are negative, which is quite different from the
quark model prediction of $x \delta d_v$ being positive for most values of $x$
\cite{Sather,RodionovTL,BeneshG,BeneshGS,BeneshL}.
Furthermore, our numerical results are about 10\% of the
quark model estimation \cite{Sather,RodionovTL,BeneshG,BeneshGS,BeneshL}.
It has been argued
that although the absolute values of $\delta d_v$ and $\delta u_v$ are small,
the ratio $R_{\rm min}=\delta d_v/d^p_v$ may be much larger than the
ratio $R_{\rm maj}=\delta u_v/u^p_v$ in the large-$x$ region
since the $d^p_v(x)/u^p_v(x) \ll 1/2$ as $x \ra 0$,
and values as large as $5\% \sim 10\%$ \cite{Sather,RodionovTL,BeneshG,BeneshGS,BeneshL}
have been obtained for the ratio $\delta d_v/d^p_v$.
No such large-$x$ enhancement appears in our calculation for both ratios.
We find that the ratio $\delta d_v/d^p_v$ exhibits a maximum about
$0.2\%$ at $x=0.1$ while the ratio $\delta u_v/u^p_v$
diverges as $x \ra 0$ but is smaller than $0.3\%$ in the region of $x > 0.02$.
The numerical results for the CSV in the sea quark
($x \delta \dbar$, $x \delta \ubar$, $x \delta s$ and $x \delta \sbar$)
are given Fig.~5.
We find that $x \delta \dbar$ has the largest CSV and that 
$x\delta d_v$ and $x \delta u_v$ are of similar magnitude, which is consistent with
our expectation from the analysis of the fluctuation functions.
Our prediction for the $x \delta \dbar$ being negative
is opposite to the positive theoretical prediction in \cite{BeneshL}.
Our calculation for the low-$x$ behaviours of 
$x \delta \dbar$, $x \delta \ubar$, $x \delta s$ and $x \delta \sbar$
are also quite different from the quark model prediction \cite{BeneshL} -- 
the quark model predicts that these quantities diverge as $x \ra 0$ while our 
calculations show all these CSV distributions go to $0$ as $x \ra 0$.
We did not find any significant large CSV in the sea quark distribution of the
nucleon, which is consistent with the most recent phenomenological
analysis \cite{BorosSLT}.
We would like to emphasise that instead of the quark model
we adopt a totally different model, the meson cloud model, to calculate
the charge symmetry violation in the parton distributions of the nucleon.
The quark model calculation in the small-$x$ region is not very reliable
since the quark-diquark picture that is employed breaks down in this region.
The meson cloud model is suitable in the study of the CSV in the parton
distribution of the nucleon since it has the same non-perturbative origin
as the $\dbar/\ubar$ asymmetry in the proton.

\section{Criticism of Quark Model Calculations}

We have already mentioned a few of the difficulties with CSV calculations using 
quark models. A recent paper by Benesh and Londergan \cite{BeneshL} 
attempted to avoid any quark model specifics and relate possible CSV in the 
valence quark distributions to the measured valence distributions. 
Starting from the parton model expression for a quark distribution \cite{Jaffe}
\bea
q(x)  & = & 
p^+ \sum_n \delta(p^+(1-x) - p^{+}_{n}) \left| \langle n | \Psi(0) | p \rangle \right|^2
\label{eq:parton}
\eea
where the intermediate state $|n \rangle$ has 4-momentum $p_n$ and the 
+ components of momenta are defined by $k^+ = k^0 + k^z$, and then making 
the assumption that the intermediate state can be modelled by a diquark system 
with definite mass $M_d$, Benesh and Londergan investigate the consequences 
for CSV of varying $M_d$. 
Following the Adelaide group \cite{SchreiberST}, 
we can attempt to determine the dependence of the quark distribution on $M_d$. 
Assuming that the modulus squared of the wavefunction for the 
struck quark in the nucleon is symmetric about the z-axis, we can use the 
delta function to perform the integration over transverse diquark momenta
\bea
\int d{\bf p}_n \delta(p^+(1-x) - p^{+}_{n}) & = & 
2 \pi \int_{p_{min}}^{\infty} dp_n p_n
\eea
where 
\bea
p_{min} & = & \left| \frac{M^2(1-x)^2 - M_{d}^{2}}{2M(1-x)} \right| \\
p_{T} & = & 2M(1-x) \sqrt{M_{d}^{2} + {\bf p}_{n}^{2}} - M^2(1-x)^2 - M_{d}^{2}
\eea
and $M$ is the nucleon mass.
Therefore we obtain the quark distribution in the form
\bea
q(x) & = & \int_{p_{min}(x, M_{d})}^{\infty} dp_n g(p_n)
\eea
where $g(k)$  only depends on the magnitude of the 3-momentum (this is 
not true in the case of spin dependent quark distributions), and we have reminded 
ourselves that $p_{min}$ is a function of $x$ and $M_d$.
Thus all the $M_d$ dependence of the quark distribution is in the lower limit of 
the integral. Use of the fundamental theorem of calculus then gives
\bea
\frac{\partial q(x)}{\partial M_d} & = & 
\frac{\partial q(x)}{\partial x} \frac{\partial p_{min}}{\partial M_d} / 
\frac{\partial p_{min}}{\partial x} \\
& = & \frac{2M_{d}(1-x)}{M^2(1-x)^2 +M_{d}^2 } \:
\frac{\partial q(x)}{\partial x}.
\label{eq:deltaq}
\eea
This expression is similar to that of Benesh and Londergan \cite{BeneshL}, 
except that in their case the $\partial/\partial x$ operator acts on the product 
of the kinematic factor $2M_{d}(1-x)/(M^2(1-x)^2 +M_{d}^2 )$ 
and the original quark distribution. 
The derivation of reference \cite{BeneshL} 
differs from ours in that they make a variation in $M_d$ under the integral 
in equation (\ref{eq:parton}), 
then evaluate the integral over ${\bf p}_n$ by ignoring any transverse 
momenta of the diquark.
However in our expression, all transverse momenta have been properly 
integrated over (in the parton model the transverse momentum of the struck 
quark vanishes).

Benesh and Londergan then use the idea of Close and Thomas \cite{CloseT}
that the quark model $SU(4)$ spin-isospin symmetry is broken by the color 
hyperfine interaction. 
The hyperfine interaction leads to a splitting in the masses of the spin-0 and 
spin-1 diquark states and hence to a difference between the up and down 
valence distributions:
\bea
u_{v}(x) & = & \frac{3}{2} q_{v}^{s}(x) + \frac{1}{2} q_{v}^{t}(x) \nonumber \\
d_{v}(x) & = & q_{v}^{t}(x)
\eea
where the superscripts $s, t$ refer to singlet and triplet diquark states 
respectively.
If the $N-\Delta$ mass splitting is also caused by the color hyperfine 
interaction, then the shifts in the singlet and triplet diquark masses are 
found to be 
\bea
\delta_{hf}M_{d}^{t} & = & -\frac{1}{3} \delta_{hf}M_{d}^{s} = + 50 {\rm MeV}.
\eea
By now expanding $q_{v}^{s}(x, M_d)$ and $q_{v}^{t}(x, M_d)$ in Taylor series 
to first order in $\delta M_d$ about the symmetry point 
$M_d = M_{d}^{0}, \;  q_{v}^{s}(x, M_{d}^{0}) = q_{v}^{t}(x, M_{d}^{0})$
Benesh and Londergan obtain the shift in the triplet quark distribution
\bea
\delta_{hf} q_{v}^{t}(x) & = & \frac{1}{6} (2 d_{v}(x) - u_{v}(x)).
\eea
Now as the CSV at the quark level comes from quark mass and electromagnetic 
effects, both of which are iso-vector, the only mass shift is in $M_{d}^{t}$, and
to first order the shift in the triplet quark distribution will be proportional to that 
from the hyperfine interaction
\bea
\delta_{CSV}  q_{v}^{t}(x) & = & \frac{\delta_{CSV}M_{d}^{t}}{\delta_{hf}M_{d}^{t}} \:
\frac{2 d_{v}(x) - u_{v}(x)}{6}.
\eea

The main difficulty with this argument is that it is entirely based on first order 
shifts in the quark distributions. 
However the second order terms can be estimated, and they are of similar 
magnitude to the first order terms. 
Expanding the quark distributions to second order in $\delta M_d$ about the 
symmetry point we have
\bea
q_v(x, M_{d}^{0} + \delta M_d) & = & q_v(x, M_{d}^{0}) + 
\delta M_d \: \frac{\partial q_v(x, M_d)}{\partial M_d} \left |_{M_d = M_{d}^{0}} \right. + 
\frac{1}{2} (\delta M_d)^2 \: \frac{\partial^2 q_v(x, M_d)}{\partial (M_d)^2} \left |_{M_d = M_{d}^{0}} \right. 
\eea
where the partial derivatives on the right hand side can be evaluated 
using (\ref{eq:deltaq}).
This then gives for the hyperfine shift in the triplet quark distribution
\bea
\delta_{hf} q_{v}^{t}(x) & = & -\frac{1}{3} \delta_{hf} q_{v}^{s}(x) + 
2 (\delta M_{d}^{t})^2 \frac{\partial^2 q_v(x, M_d)}{\partial (M_d)^2} \left |_{M_d = M_{d}^{0}} \right. 
\nonumber \\
& = & \frac{1}{6} (2 d_{v}(x) - u_{v}(x)) + 
\frac{3}{2} (\delta M_{d}^{t})^2 \frac{\partial^2 q_v(x, M_d)}{\partial (M_d)^2} \left |_{M_d = M_{d}^{0}} \right. .
\eea
As an estimate of the second order term we can approximate $q_v(x, M_{d}^{0})$ 
by $d_v{x}$ or $(u_v(x) + d_v(x))/3$ and use one of the well-known 
parametrizations of the quark valence distributions \cite{GRV98,MRSTCTEQ}. 
We also take our value for $M_{d}^{0}$ to lie in the range $(0.65 - 0.85) M$, 
which is the range suggested by quark models, though the results are not very 
sensitive to the value of $M_{d}^{0}$ in this range.
In Fig. 6 we compare the first and second order terms for 
$\delta_{hf} q_{v}^{t}(x)$. 
We can see that our estimate of the second order term is of comparable 
magnitude to the first order term over most of the $x$-range. 
Indeed for low $x$ it is larger, showing the unreliability of quark model calculations 
in this region. 
For $x > 0.2$ the second order term is of opposite sign to the first order term, 
which indicates that the first order estimate of the shift in the quark 
distribution is too large in this region. 
This in turn implies that the estimate of CSV induced shifts in the quark 
distributions in this region are also too large.
These conclusions are not much influenced by the choice of valence quark 
parametrization, the value chosen for $M_{d}^{0}$, or whether we use Benesh and 
Londergan's expression for the dependence of the quark distribution on $M_d$ 
rather than equation (\ref{eq:deltaq}). 
The reason for these conclusions not being greatly influenced by the choice of 
expression for $\partial q(x) / \partial M_d$ is that, with the quark distributions 
used, the highest order derivative term in $x$ always dominates. 
This is a consequence of the divergences in the valence quark distributions near 
$x = 0$, $q_v(x) \sim x^{-0.5}$ in all cases.

\section{Summary}

Although it has been generally assumed that charge symmetry 
was highly respected in the nucleon system,
there have been some phenomenological analysis \cite{BorosLT,BorosSLT}
and theoretical calculations
\cite{Sather,RodionovTL,BeneshG,BeneshGS,BeneshL,Ma}
about the possible extent of CSV in the parton distributions.
Any unexpected large CSV will greatly affect our understanding
on the non-perturbative dynamics and hadronic structure,
and the extraction of $sin^2\theta_W$ from neutrino scattering.
Up to now most theoretical attempts to calculate the CSV
in the parton distributions are based on the quark model
and employ the quark-diquark model. 
In this paper we point out that CSV in both the valence and sea quark
distributions of the nucleon can arise from the non-perturbative dynamics
of the nucleon.
We present an alternative analysis of CSV in the parton
distributions employing the meson cloud model,
which has previously been successful in the study of the flavour asymmetry
and the quark-antiquark asymmetry of the nucleon.
In the meson cloud model the proton and neutron may fluctuate
into hadron-meson Fock states in which
the hadrons and mesons are in different charged states respectively.
As we consider the mass differences among these hadrons
and mesons, the probabilities of proton and neutron
fluctuating into the corresponding Fock states will be different.
Thus the non-perturbative contributions to the valence and sea quarks
distributions will be different, which naturally leads to the CSV in both the
valence and sea distributions of the nucleon.
Our predictions for the CSV in the valence sector and sea sector
are both different from the quark model calculations. 
We also point out the deficiencies of quark model based calculations
of CSV in the parton distributions. 
In particular the quark-diquark picture is inadequate at low-$x$, and in 
the medium-$x$ region the use of a first order shift in the parton distributions 
must be questioned, as higher order shifts are of similar magnitude.
The coming experimental information on the parton distributions of the nucleon
and more theoretical studies on this issue will
examine these calculations.

\section*{Acknowledgments}
This work was partially supported by the Massey Postdoctoral Foundation,
New Zealand.
We are grateful to Prof. A. W. Thomas for useful discussions.
F.~G. Cao would like to thank
the Special Research Center for the Subatomic Structure of Matter for its hospitality.

\newpage
\begin{center}
{ Table 1. Parameters in Eq.~(\ref{qfit})} at $Q^2=4$ GeV$^2$.

\vskip 0.5cm
\begin{tabular}{|c|c|c|c|c|c|}\hline
   & $a$ & $b$ & $c$ & $d$ & $e$ \\ \hline
$v^\pi(x,Q^2)$ & $1.712$ & $-0.518$ & $1.182$ & $-0.836$ & $0.972$ \\ \hline
$\sbar^K(x,Q^2)$ & $0.803$ & $-0.516$ & $1.306$ & $-0.762$ & $0.957$ \\ \hline
$u^p(x,Q^2)$ & $1.029$ & $-0.572$ & $3.933$ & $1.550$ & $6.033$ \\ \hline
$d^p(x,Q^2)$ & $0.615$ & $-0.575$ & $5.096$ & $1.102$ & $6.773$ \\ \hline
\end{tabular}
\end{center}

\newpage
\section*{Figure Captions}
\begin{description}
\item
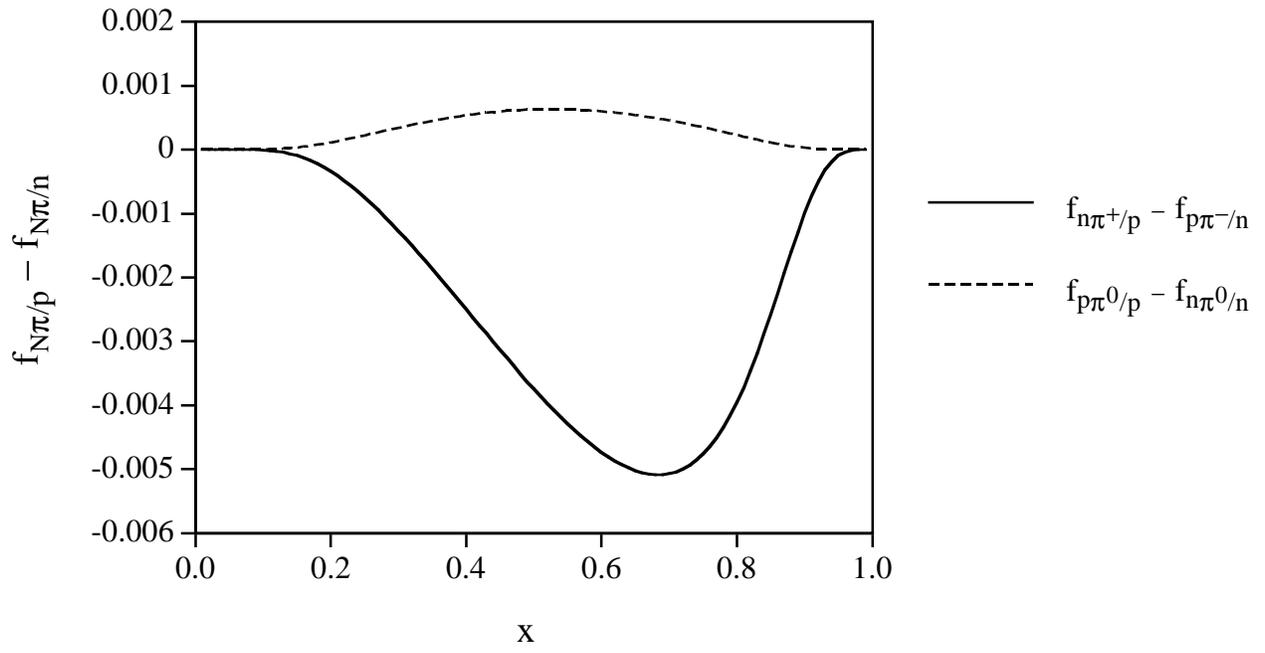
{Fig.~1.} 
The differences between fluctuation functions
$f_{N\pi/p}$ and $f_{N\pi/n}$.
\item
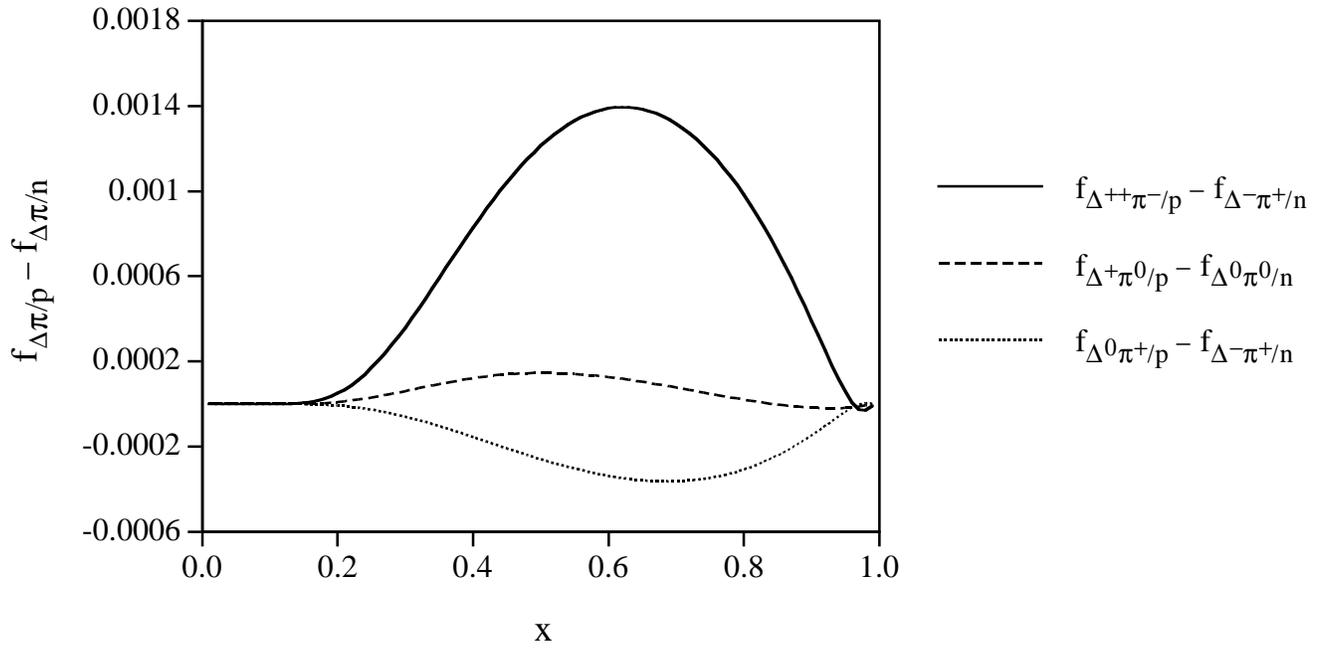
{Fig.~2.}
The differences between fluctuation functions
$f_{\Delta\pi/p}$ and $f_{\Delta\pi/n}$.
\item
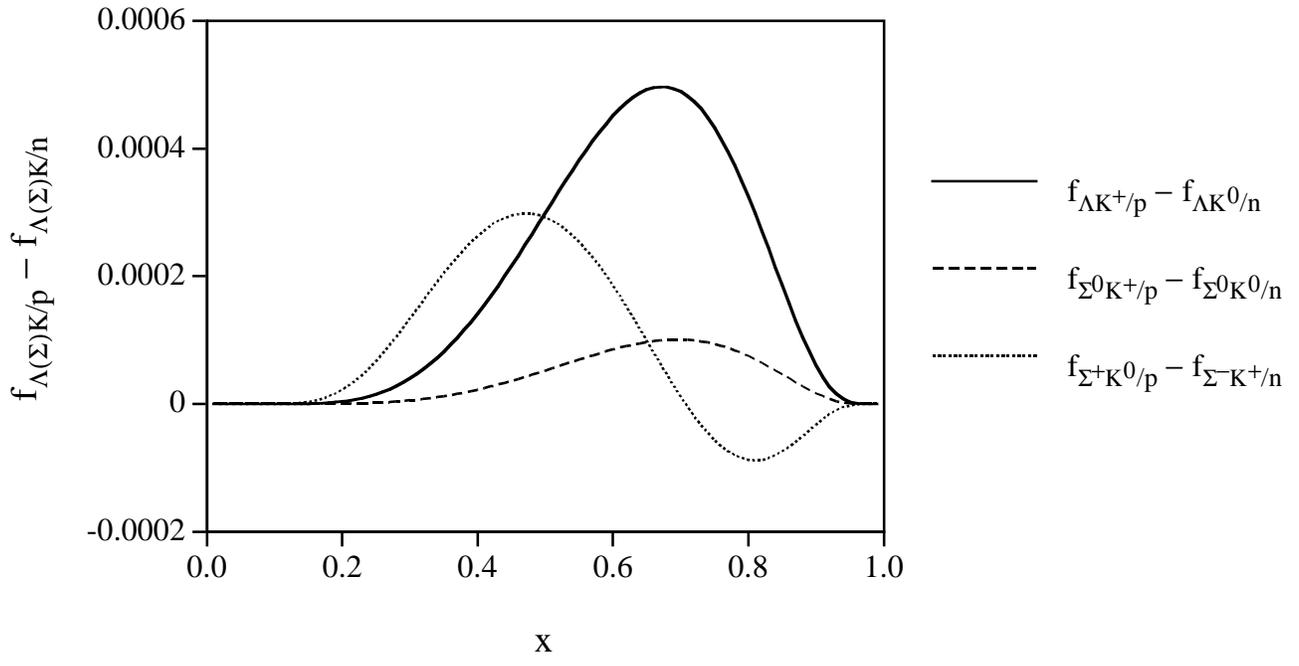
{Fig.~3.}
The differences between fluctuation functions
$f_{\Lambda(\Sigma)K/p}$ and $f_{\Lambda(\Sigma)K/n}$.
\item
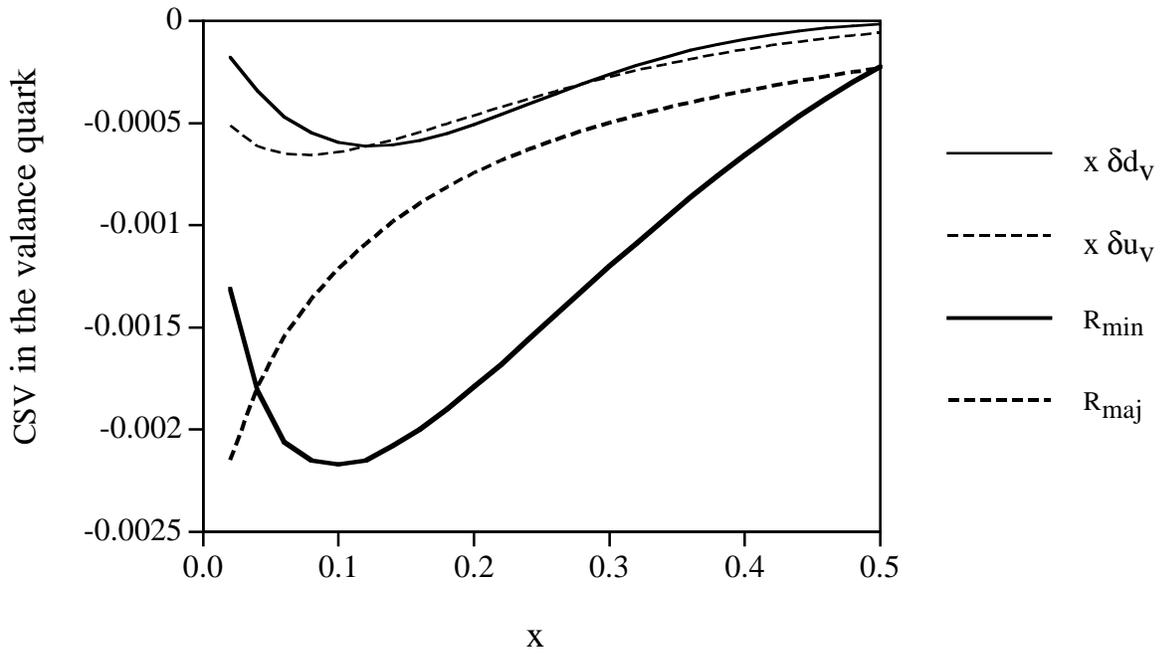
{Fig.~4}
The charge symmetry violation in the valence quark sector.
$R_{\rm min}=\delta d_v/d^p_v$ and
$R_{\rm maj}=\delta u_v/u^p_v$.
\item
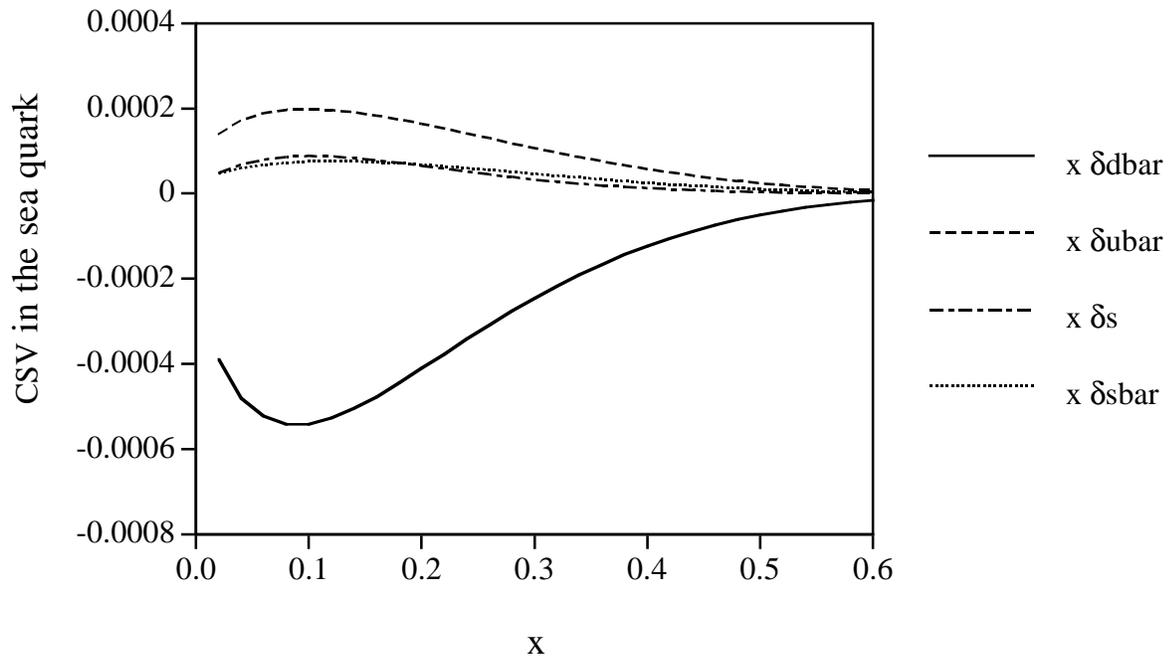
{Fig.~5}
The charge symmetry violation in the sea quark sector.
\item
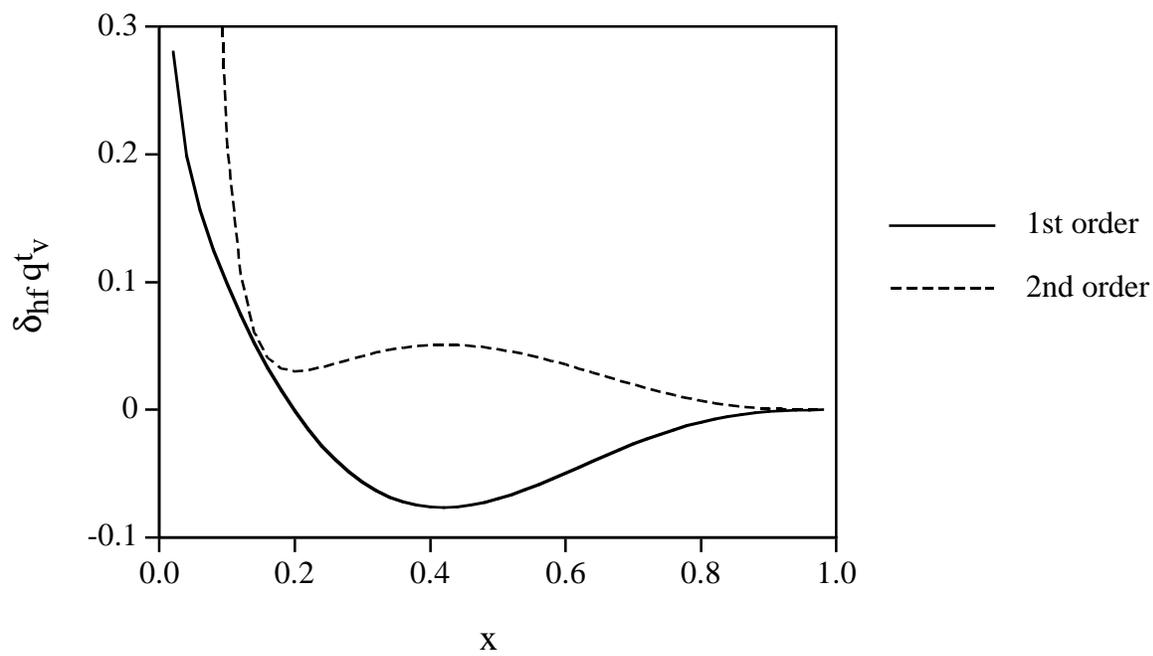
{Fig.~6}
The first and second order shifts in the triplet quark distribution 
caused by the color hyperfine interaction. 
The solid curve is the first order shift $(2d_v(x) - u_v(x))/6$ calculated 
using the parametrizations of reference \cite{GRV98}. 
The dashed curve is the second order shift estimated using 
$q(x) =  (d_v(x) + u_v(x))/3$, $M_d^0=0.75 M$,
and a mass shift of 50 MeV for the triplet diquark state.
\end{description}

\end{document}